\documentclass[12pt]{iopart}
\usepackage{graphicx}
\newcommand{\TMYAG}{Tm$^{3+}$:YAG }
\newcommand{\NDYVO}{Nd$^{3+}$:YVO$_{4}$ }
\begin{document}
\title{Efficient light storage in a crystal using an Atomic Frequency Comb}
\author{T. Chaneli\`ere$^1$, J. Ruggiero$^1$, M. Bonarota$^1$, M. Afzelius$^2$, J.-L. Le~Gou\"et$^1$}
\address{$^1$ Laboratoire Aim\'e Cotton, CNRS-UPR 3321, Univ. Paris-Sud, B\^at. 505, 91405 Orsay cedex, France}
\address{$^2$Group of Applied Physics, University of Geneva, CH-1211 Geneva 4, Switzerland}
\ead{thierry.chaneliere@u-psud.fr}

\begin{abstract}
We demonstrate efficient and reversible mapping of a light field onto a thulium-doped crystal using an atomic frequency comb (AFC). Thanks to an accurate spectral preparation of the sample, we reach an efficiency of $9\%$. Our interpretation of the data is based on an original spectral analysis of the AFC. By independently measuring the absorption spectrum, we show that the efficiency is both limited by the available optical thickness and the preparation procedure at large absorption depth for a given bandwidth. The experiment is repeated with less than one photon per pulse and single photon counting detectors. We clearly observe that the AFC protocol is compatible with the noise level required for weak quantum field storage.
\end{abstract}
\maketitle

\section{Introduction}
The ultimate control of a quantum light field naturally demands a good understanding of its interaction with matter. In that context the reversible mapping of a field into an atomic system represents a fundamental challenge. The experimental quest of a quantum memory for light has produced many accomplishments using different systems. This domain has also been motivated by the realization of a quantum repeater for which the memory is a critical element \cite{qrep}. Benefitting from a long experience, most of the recent achievements have involved atomic vapors \cite{Julsgaard, Eisaman, Chaneliere, Choi, Chen}. More recently atom-like impurities in solid state matrices have been considered as excellent candidates for a quantum interface. They can indeed have good coherence properties \cite{Atature, jelezko}. One can take advantage of their complete absence of motion to demonstrate extremely long coherence time \cite{fraval, longdell, louchet}. The rare-earth ion doped crystals (REIC) are particularly interesting because they are relatively easy to produce and commercially largely available. The most promising storage protocols fully use the primary feature of the REIC: the optical transition have a long coherence lifetime and a large inhomogeneous broadening. These schemes are in the lineage of the photon echo technique and are intimately related to a dipole rephasing. The direct excitation of the optical coherences over a wide spectral range (larger than the homogeneous width) is usually the initial storage step. The  role of the protocol is to rephase the dipoles by externally controlling the detuning of the transition \cite{crib1, alexander2006, crib3, hetet}, which can lead to large storage efficiencies \cite{tittel}.

More recently a protocol based on the preparation of an Atomic Frequency Comb (AFC) has allowed light storage at the single photon level \cite{afzelius, AFCnature}. The protocol is based on the absorption of a photon by a spectrally periodic structure of narrow peaks (a comb of periodicity $1/T$). Right after absorption, each spectral class of dipoles starts dephasing with respect to each other. But at the time $T$, the accumulated phases are all equal to first order to an integer number of 2$\pi$ (0 modulo 2$\pi$): this coherence build-up triggers a retrieval. Using a temporal image, this procedure can be interpreted as a dipole rephasing. Alternatively a spectral image can also be employed. The periodic spectral filter is a diffraction grating. It produces a delayed response in the time domain (the retrieval). This spectral image has been widely used to explain the stimulated photon echo (SPE). The AFC is in the direct lineage of the SPE where a periodic modulation of the absorption profile is considered. In that sense, the AFC represents an extreme situation: high narrow absorbing peaks are separated by fully transparent region. Such a comb can absorb close to 100\% of the light and reduces intrinsic dephasing during the storage time $1/T$. As a consequence the retrieval efficiency should be high in the forward direction (54\%) and ideally perfect (100\%) in the backward configuration \cite{afzelius}. In order to achieve high efficiency, it is thus crucial to achieve a precise spectral shaping of the inhomogeneous absorption profile. The AFC is not only an interesting alternative storage protocol. It has been shown to have the highest temporal multimode capacity in principle \cite{nunn}, which is a critical figure of merit in the prospect of quantum communication \cite{Christoph}.

In this paper we show how the AFC protocol can be efficiently implemented in an appropriate crystal namely \TMYAG. Our work is focused on the initial step of the protocol where the signal is stored onto the optical coherences. Our results in terms of storage efficiency  correspond to more than an order of magnitude improvement as compared to previous proof of principle realization performed in \NDYVO \cite{AFCnature}. This improvement is possible  because the Tm:YAG allows for a better preparation of the AFC due to an efficient optical pumping and a narrower homogeneous linewidth. More generally the material properties gives the ultimate width of the spectral selection (presence of spectral diffusion, superhyperfine effect ...). Additionally any external source of broadening as the laser linewidth can reduce the efficiency for a given bandwidth. This limiting factor is not present in our case since we use a stabilized laser. 

\section{Efficient light storage in \TMYAG}
Thulium has an interaction wavelength (793nm) easily accessible with laser diode as compared to other rare-earth ions (praseodynium and europium). This is particularly convenient when a complex spectral preparation sequence is required. The trivalent thulium is also known to have a long optical coherence time from 10 to 100 $\mathrm{\mu s}$ depending on the crystalline matrix, dopant concentration and magnetic field \cite{macfarlane}. The spin coherence lifetime has been also observed to be long in the ground state \cite{louchet}. Very precise spectral tailoring will ultimately lead to a high number of temporal modes stored in the medium, which primarily scales as the number of spectral channels accessible in the absorption profile. The long hole-burning lifetime observed under an appropriate magnetic field will be especially useful for the initial state preparation \cite{ohlsson}. This ensures an efficient population transfer between hyperfine levels (Zeeman states) by optical pumping \cite{Lauritzen}. 

\subsection{Experimental setup}
\begin{figure}[ht]
\centering
\includegraphics[width=12cm]{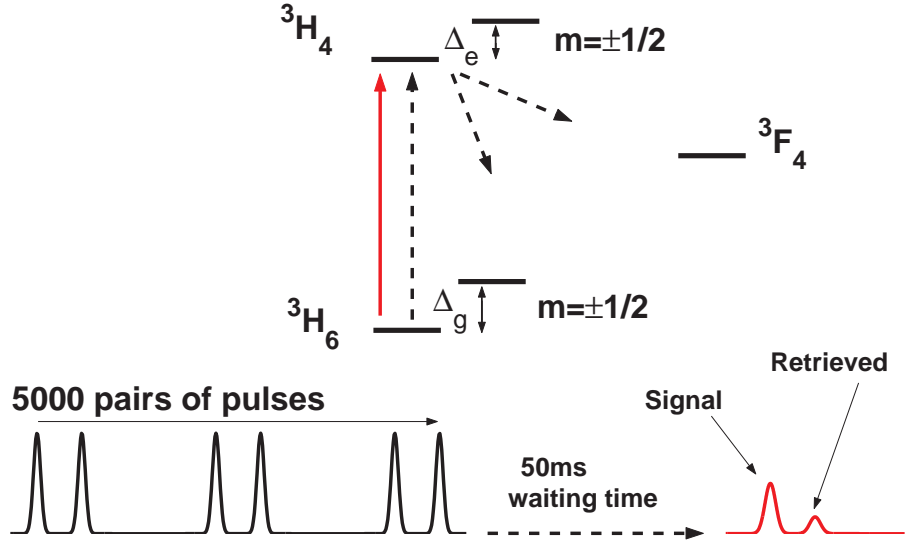}\caption{Relevant level structure of the trivalent thulium under magnetic field. The optical pumping between the two Zeeman states is mainly allowed by a weak branching through the $^{3}\textrm{H}_{4}(0)$ excited state. We also represent schematically the pulse sequence (see text for details).}
\label{fig:FIGtimeseq}
\end{figure}

To implement the AFC protocol in \TMYAG, we apply a 210G magnetic field along the [001] crystalline axis. This splits the ground and excited levels into a nuclear spin doublet by $\Delta_g=6 $MHZ and $\Delta_e=1.3$ MHz respectively \cite{OGN} (see fig. \ref{fig:FIGtimeseq}). This is sufficient to create a few megahertz interaction bandwidth ($\sim$ 3 MHz in our case). Our 0.5\% doped- \TMYAG crystal is immersed in liquid helium at 2.3K. This permits a good coherence lifetime ($T_2=30 \mu s$) and then an accurate spectral tailoring. This is also required to obtain a long lived periodic structure in the ground state: 7s in our case critically depending on the temperature \cite{ohlsson}. The AFC structure is built by optical pumping between the spin doublets. Although small (typically 2\% for our field orientation \cite{OGN}), the optical branching ratio is compensated by a long lifetime of the ground state population. This is suffiecient to efficiently pump the ions between the long-lived spin levels. The laser polarization is also applied along the [001] axis. In this configuration the splitting is the same for the different crystallographic sites interacting with the laser. The bandwidth of the AFC is actually larger than $\Delta_e$ because we use a moderate magnetic field. This could be an issue in practice because one would obtain the interference of two AFCs. In order to avoid this effect, we simply make the comb period $1/T$ coincide with the excited level splitting ($\Delta_e=2/T$ in our case).

The spectral preparation is done by applying a sequence of pulse pairs (FWHM duration $\sim 300 \mathrm{ns}$ separated by $T=1.5 \mathrm{\mu s}$ and then followed by a $100 \mathrm{\mu s}$ dead time before the next pair). This train consists of 5000 pairs followed by a long waiting time ($50 \mathrm{ms}$) to ensure the complete decay of population from the $^{3}\textrm{H}_{4}(0)$ excited state (lifetime $\sim 800 \mathrm{\mu s}$). Each pair whose area is much smaller than $\pi$ creates a small population difference \cite{AFCnature} with a periodicity of 1/T in frequency space. After accumulation, the sequence gives rise to the atomic frequency comb. To first probe the frequency spectrum and to later generate a pulse to be stored, we use a very weak independent signal beam.
Temporal shaping and frequency scanning is achieved by two independent acousto-optic modulators (AOM). The pumping beam has been also designed to be twice as large as the signal in the crystal. The AFC is then spatially uniform over the entire signal beam.

\subsection{Atomic Frequency Comb spectrum and storage efficiency}

\begin{figure}[ht]
\centering
\includegraphics[width=12cm]{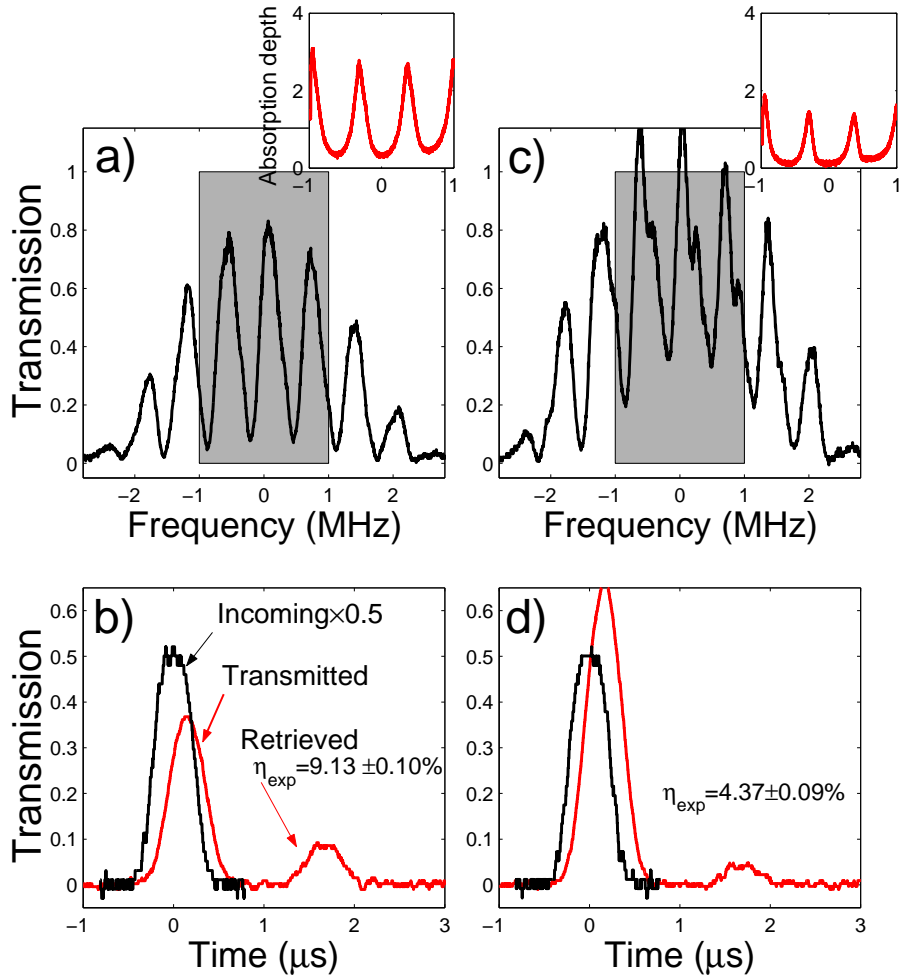}\caption{a) and c): AFC prepared for different pumping powers. The 660 kHz period is given by the delay between the preparation pulses $T$. The insets shows the central part of the optical depth spectrum.b) and d): A weak signal is sent matching the AFC bandwidth of a) and c) respectively. After being partially absorbed, it gives rise to a retrieval. The efficiency strongly depends of the shape of comb, which in turn depend on the power of the preparation pulses: the power is larger for the set c)-d) than for a)-c).}
\label{fig:FigAFC_article}
\end{figure}

By sweeping the signal AOM driving frequency, we can directly observe the transmission spectrum (see fig. \ref{fig:FigAFC_article}a and \ref{fig:FigAFC_article}c). Even if the pumping power is very weak, we see that the AFC structure depends strongly on the preparation pulse intensity. For an increasing power from \ref{fig:FigAFC_article}a to \ref{fig:FigAFC_article}c, the peak maximum absorption decreases due to power broadening of the hole-burning process (see insets). The sweep rate $\left(320\mathrm{kHz}\right)^2$ being slightly too high, we observe a distortion of the observed spectrum (see fig. \ref{fig:FigAFC_article}c for example). This effect is not apparent in the insets representing only the central part of the spectrum where a slower sweep rate $\left(120\mathrm{kHz}\right)^2$ has been used.

A weak pulse to be stored is then sent onto the medium. To be sure that the signal is uniformly covered by the comb, its duration is chosen to be slightly longer than the preparation pulses (FWHM $\sim 450 \mathrm{ns}$). 
When a signal is absorbed into the medium, it produces a retrieval. Curve \ref{fig:FigAFC_article}b (resp. \ref{fig:FigAFC_article}d) corresponds to \ref{fig:FigAFC_article}a (resp. \ref{fig:FigAFC_article}c). To be able to compare the incoming pulse and the retrieved signal, we simply bleach the medium by applying a strong single pumping beam (stronger than the preparation pulses). This procedure gives a reference for a fully transmitted incoming pulse. We can then estimate the total efficiency: $\eta_{exp}=9.13\pm0.10\%$ (resp. $\eta_{exp}=4.37\pm0.09\%$) for the comb \ref{fig:FigAFC_article}b (resp. for \ref{fig:FigAFC_article}d). This corresponds to more than one order of magnitude improvement as compared to the previous proof of principle realization \cite{AFCnature}.

\section{Discussion}

We now investigate the storage efficiency thanks to the independently measured AFC spectrum. To do so, we have developed an original model to analyze our data. The expected efficiency has already been calculated by assuming that the comb is composed of well-separated gaussian peaks \cite{afzelius}. This calculation was based on a temporal image by following the evolution of the atomic variables during the time sequence. Here we derive a more general formula valid for any peak shape.

\subsection{Expected efficiency}

We consider the propagation of a field $\mathcal{E}$ in the spectral domain:
$$\displaystyle
\partial_z^2\mathcal{E}(z,\omega)+k^2\left(1+\chi(\omega)\right)\mathcal{E}(z,\omega)=0
$$
The action of the medium is represented by the susceptibility $\chi(\omega)$ ($k$ is the wave vector). To account for the periodic structure of the AFC (period $1/T$), we naturally decompose $\chi(\omega)$ in a Fourier series: $$ \chi(\omega)=\sum_{p \geq 0} c_p \exp \left(2i\pi p \omega T\right)$$

The components corresponding to $p<0$ are dropped for causality reason. We here calculate the efficiency in the forward direction but our analysis could be extended to describe the backward retrieval \cite{afzelius}. The field $\mathcal{E}(z,\omega)$ has the same periodic structure:
$$ \mathcal{E}(z,\omega)=\mathcal{E}(0,\omega)\sum_{p \geq 0} a_p(z) \exp \left(-2i\pi p \omega T\right)
$$
Each term of the decomposition gives the amplitude of a retrieval centered at the time $pT$. The first term $p=0$ represents the absorption of the incoming signal. In the forward direction, the efficiency of the AFC protocol is then directly given by $\mid a_1\mid^2$. All the coefficients can be recursively determined by solving the propagation equation for $a_p(z)$. The first two components are given by:
$$
\begin{array}{l}
\partial_z^2 a_0(z)+k^2\left(1+c_0\right)a_0(z)=0\\[.1cm]
\partial_z^2 a_1(z)+k^2\left(1+c_0\right)a_1(z)+k^2c_1a_0(z)=0
\end{array}
$$

The efficiency in the forward direction is then given by
\begin{equation}\label{eta_c1c0}
\eta=\frac{1}{4} \frac{\mid c_1\mid^2}{\mathrm{Im}\left(c_0\right)^2} {\tilde{d}}^2 \exp\left(-\tilde{d}\right)
\end{equation}
$\tilde{d}$ is the average absorption depth defined as $\tilde{d}=-k \mathrm{Im}\left(c_0\right)L$ where $L$ is the length of the medium. Our formula is very general and just assumes that the susceptibility is periodic. It is sufficient to estimate the expected efficiency for a given frequency comb as presented in fig. \ref{fig:FigAFC_article}. Our model doesn't make any assumption on the exact shape of the profile. This is the main advantage of our model. It also represents an alternative interpretation of the phenomenon in the spectral domain: the periodic filter is considered as a diffraction grating. A simple numerical Fourier expansion of the absorption spectrum (see insets of fig. \ref{fig:FigAFC_article}a and \ref{fig:FigAFC_article}c) gives the different coefficients to determine $\eta$. We calculate $\eta=7.8\%$ (resp. $\eta=3.5\%$) for fig. \ref{fig:FigAFC_article}a (resp. \ref{fig:FigAFC_article}c). More generally for different preparation pulses intensities, we find a correct agreement between the experimental efficiencies and the ones deduced from the measured absorption spectrum ($\eta_{exp}$ is represented by a red dashed line and $\eta$ by a black dashed line in fig. \ref{fig:OptAFC_article}). Nevertheless we systematically observe that the measured values are larger than the prediction. This may be due to inaccurate measurement of the absorption spectrum. As already mentioned, the lower sweep rate $\left(120\mathrm{kHz}\right)^2$ used to obtain  the central part of the spectrum may still be too high as compared to the width of the AFC peaks. This actes as a convolution effect, artificially broadening the peaks. More generally it is not fully obvious to accurately measure an optical thickness much larger than one. Any very far off resonance contribution of the laser would be transmitted through the medium. Here it could be due to the additionnal tapered-amplifier whose fluorescent background is extremely broad. For both effects, the real optical thickness would be underestimated and so would be the predicted efficiency.

\subsection{Current limitations of the pumping scheme}

\begin{figure}[ht]
\centering
\includegraphics[width=12cm]{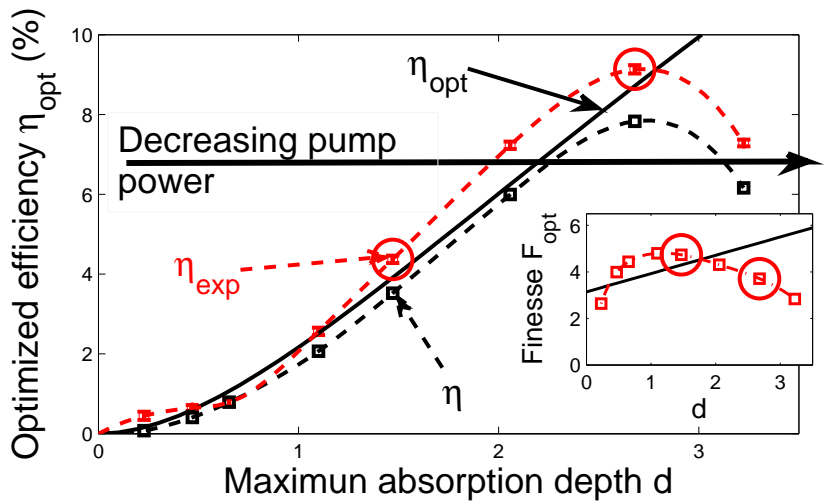}\caption{Color online: Measured efficiency $\eta_{exp}$ as a function of the maximum optical depth (red dashed line used to guide the eye). We have also calculated the expected efficiency $\eta$ (eq. \ref{eta_c1c0}) from the absorption spectrum (black dashed line used to guide the eye). Then assuming that the comb is composed of lorentzian peaks (eq. \ref{effLorentz}), we have finally plotted the optimum efficiency $\eta_{opt}$ (solid black line) and fitted the experimental spectrum to obtain the finesse. Inset: Measured finesse obtained by fitting the absorption spectrum with a comb of lorentzian peaks (red dashed line used to guide the eye) and comparison to the optimum $F_{opt}$ (solid black line). The two cases presented in fig. \ref{fig:FigAFC_article} are circled.}
\label{fig:OptAFC_article}
\end{figure}

It is now important to understand the current limitations. We will first estimate the maximum efficiency that could be obtain in our case and compare it to our measurements (see fig. \ref{fig:OptAFC_article}). The total efficiency strongly depends on the exact shape of the frequency comb. For a given initial optical thickness, one has to properly choose the peak width to optimize the retrieval. We now assume that the comb is composed of lorentzian (half width at half maximum $\Gamma$) periodically distributed peaks. The lorentzian shape allows simple analytical calculation and gives a satisfying agreement with the measured spectrum. The efficiency depends only on the maximum optical depth $d$ and the comb finesse $F=\pi/\left(\Gamma T\right)$ \cite{afzelius}.
\begin{equation}\label{effLorentz}
\eta=d^2 \tanh^2\left(\pi/2F\right) e^{-d \tanh\left(\pi/2F \right)}e^{-2\pi/F}
\end{equation}

This result can be compared to eq. 1 of \cite{AFCnature}. The structure and the term interpretation are similar. Our formula includes by definition a potential overlap of the lorentzian peaks when the finesse is low. This has been alternatively interpreted by de Riedmatten \textit{et al.} as a uniform absorbing background and fitted independently \cite{AFCnature}. In our case, we can directly fit the AFC spectrum with eq. \ref{effLorentz}.
Our result can be also simplified when the finesse is high. In that case, the best efficiency of the protocol is expected. For a given initial optical thickness $d$, the finesse should be properly chosen to obtain a optimized efficiency. For $F_{opt}=\pi \left( 1+d/4 \right)$, we have
\begin{equation}\label{effopt}
\eta_{opt}=4e^{-2} d^2/\left({4+d}\right)^2
\end{equation}
The efficiency is limited to 54\% in the forward direction. This value is a common feature of the protocols involving a dipole rephasing \cite{hetet}.

One can now wonder if the measurements correspond to the optimum for a given optical depth. As we have seen, the shape of the comb is essentially controlled by varying the intensity of the preparation pulses. At the same time this procedure is changing the maximum absorption depth $d$ and the width of the peaks (the finesse $F$). For different pumping power, we can directly measure the maximum absorption depth $d$ and then estimate the corresponding $\eta_{opt}$ (eq. \ref{effopt}).

The comparison between the measured efficiencies $\eta_{exp}$ (red dashed line in fig.\ref{fig:OptAFC_article}) and the expected optimum for a given optical depth $\eta_{opt}$ (solid black line in fig.\ref{fig:OptAFC_article}) allows us to evaluate the quality of the preparation. There is clearly a good matching of the efficiency curves at low optical depth (between 0 and 3). In that case, the efficiency is not critically depending on the finesse (eq. \ref{effLorentz}) \cite{afzelius}. In practice any finesse between $\pi$ and $2\pi$ could be considered as optimal. This is verified experimentally and explains the efficiency curve matching even if the finesse is only roughly following $F_{opt}$ (see inset in fig.\ref{fig:OptAFC_article}). On the contrary, we observe an efficiency breakdown at larger optical thickness ($d \sim 3$ in fig.\ref{fig:OptAFC_article}) where $\eta$ is now critically depending on the finesse (eq. \ref{effLorentz}). The measured finesse is indeed significantly deviating from $F_{opt}$ (see inset in fig.\ref{fig:OptAFC_article}). This is sufficient to explain the efficiency breakdown in this region.

This clearly shows that the preparation sequence may not be appropriate at large optical thickness. Even if we should be able to benefit from the initial optical depth of our crystal ($d \sim 4-5$), we cannot properly prepare the AFC to obtain the maximum efficiency. In practice the external control of the preparation is given by the power of the pumping pulses. This single degree of freedom cannot independently master the two parameters of the comb (height $d$ and finesse $F$). Our comparison shows that a simple pumping strategy using a train of pulse pairs is relatively appropriate at low optical thickness as soon as the power is properly adjusted. On the contrary, at large optical thickness, the efficiency is limited. 

The pumping intensity primarily affects the width of hole-burning process because of the power broadening. This effect is intimately related to the optical pumping dynamics. To break this limitation, it is possible to apply a coherent population transfer scheme \cite{roos, DESEZE05, Stirap}. This approach has been successfully implemented to obtain and isolate narrow feature in different systems including \TMYAG \cite{Rippe, DESEZE05, Stirap}. In other words, by enriching the preparation sequence with numerous pulses, one would obtain another degree of freedom to control accurately the comb shape. The propagation of sophisticated pulse sequence in highly absorbing materials may nevertheless limit the capability of the coherent population transfer. In that case, one can also use a transverse illumination of the slab \cite{side}. Combining this two techniques, it should be possible to independently control the efficiency and the spectral properties of the AFC.

\section{Storage at the single photon level}

Even if the efficiency is a critical parameter for a quantum storage protocol, a noise estimation should also be done. In order to investigate this level in the very weak field condition, we reproduce the experiment with less than one photon per pulse.

\begin{figure}[ht]
\centering
\includegraphics[width=11cm]{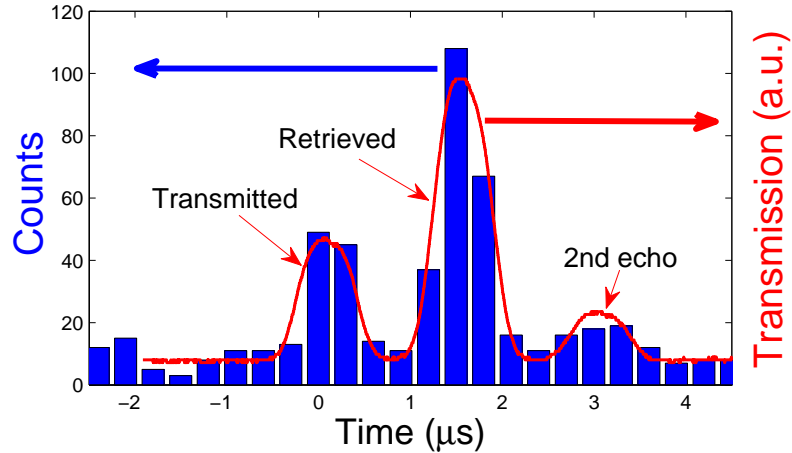}\caption{Color online: Histogram of the accumulated counts for 5.51s within a 300ns time gate. The incoming pulse contains $\sim$ 0.5 photon/pulse. We have recorded independently the signal at higher power using an standard photon diode (solid red curve). The three observed pulses are respectively the transmitted power, the retrieved signal (first echo) and a second echo (see text for details).}
\label{fig:FigureSingleC_article}
\end{figure}

We use the same preparation procedure that have been optimized previously. The weak incoming pulse to be stored is strongly attenuated by using neutral density filters. After transmission through the crystal, the signal is now fiber-coupled to a single photon counting module (Perkin-Elmer SPCM AQR 14 FC). To increase the measurement statistics, we repeat 1000 signal pulses after each preparation procedure (5000 pairs followed by a $50 \mathrm{ms}$ waiting time). The total duration of the sequence is now 329ms corresponding to 3039 signal pulses per second. The arrival of the photon is counted in a 300ns time gate and accumulated for 5.51 seconds. The corresponding histogram is presented in fig. \ref{fig:FigureSingleC_article} for different position of the time gate covering the arrival of the transmitted signal and the AFC echo.

We have independently measured the collection efficiency from the crystal to the single photon counting modules. Knowing the quantum efficiency of the detector and assuming that the AFC efficiency is the same as previously measured, we estimate the incoming number of photons per pulse in the crystal to be $\sim$ 0.5 photon/pulse.

For comparison, the measurement has been also done at higher power (solid curve in fig. \ref{fig:FigureSingleC_article}). We observe the same kind of trace as previously described (see fig. \ref{fig:FigAFC_article} for example). Nevertheless we suspect the efficiency to be higher because we also observe a second echo. This can be due to the slightly different experimental conditions (magnetic field). We haven't measured the efficiency in that case but this effect is a potential route to improve the retrieval and is currently under investigation. The photon counts (histogram in fig. \ref{fig:FigureSingleC_article}) follow the same curve within the statistical fluctuations. We additionally observe a detection background of $\sim$ 8 counts. By comparing this noise level to the maximum of the retrieved signal, we obtain a signal to noise ratio of 11. The intrinsic dark counts of the detector should be of 0.31 count with this cycling, gating rate and accumulation time. We have isolated this detection background. It is due to a leak of the preparation beam. Even if we use a first AOM to temporally shape the preparation pulse and a second one right before the detector, the total isolation is not sufficient to completely remove the residual light. In the near future, we can also use a mechanical chopper since the waiting period is long (50ms). We haven't finally seen any spurious light due to the medium itself (fluorescence). The spontaneous emission is strongly reduced after a long waiting time and is also emitted in all directions. We then believe that the ultimate limitation of such a measurement are the dark counts of the detectors. In other words, the AFC storage should not add any noise to the incoming photons. This is a remarkable figure of merit that make the protocol promising for the storage of non-classical field.

\section{Conclusion}
To conclude, we have successfully applied the AFC protocol to a \TMYAG crystal whose properties are of particular interest in the prospect of quantum repeaters. We have observed efficiencies up to $9.13\pm0.10\%$. Our measurements correspond to more than an order of magnitude improvement as compared to previous realizations in \NDYVO \cite{AFCnature}. Since the material properties gives the ultimate width of the spectral selection, they strongly influence the efficiency. This comparison shows that further improvements will demand material developments. This is especially true for the future backward configuration since \TMYAG and \NDYVO have only two ground states. This configuration does not allow, at the time, optical pumping in an auxiliary level and the application of control pulses \cite{afzelius}.

We have developed an original spectral analysis. Our model directly predictd the observed efficiency for a given initial absorption spectrum that we independently measure. For a definite bandwidth and a fixed delay, the efficiency is both limited by the available optical thickness and the preparation method. In that case, due to power broadening effect, the width and the height of the comb peaks are correlated. As a consequence, the efficiency and the spectral feature of the protocol (delay and multi-mode capacity) are not independent. These characteristics are critical for quantum repeater applications. A global estimation of the ultimate performances will demand the development of original preparation methods.

We have finally shown that even for very weak pulse $\sim$ 0.5 photon/pulse, the storage itself should not add any noise to the incoming field. The efficiency and the noise feature render then the AFC particularly attractive as a quantum memory protocol.

\end{document}